\documentstyle[prb,aps]{revtex}
\begin{document}
%\draft
\input{psfig}
%\twocolumn[
%\hsize\textwidth\columnwidth\hsize\csname @twocolumnfalse\endcsname  
\title{Surface spin-flop and discommensuration transitions in
antiferromagnets}
\author{C. Micheletti,$^1$ R. B. Griffiths$^2$ and J. M. Yeomans$^3$}
\address{(1) International School for Advanced Studies (S.I.S.S.A.) - INFM,
Via Beirut 2-4, 34014 Trieste, Italy and\\
The Abdus Salam Centre for Theoretical Physics - Trieste, Italy}
\address{(2) Physics Department,
Carnegie-Mellon University, Pittsburgh, PA 15213}
\address{ (3) Theoretical Physics, Oxford University,
1 Keble Road, Oxford OX1 3NP, UK}
\date{\today}
\maketitle
\begin{abstract}
	Phase diagrams as a function of anisotropy $D$ and magnetic
field $H$ are obtained for discommensurations and surface states for
an antiferromagnet in which $H$ is parallel to the easy axis, by
modeling it using the ground states of a one-dimensional chain of
classical $XY$ spins. A surface spin-flop phase exists for all $D$,
but the interval in $H$ over which it is stable becomes extremely
small as $D$ goes to zero.  First-order transitions, separating
different surface states and ending in critical points, exist inside
the surface spin-flop region. They accumulate at a field $H'$
(depending on $D$) significantly less than the value $H_{SF}$ for a
bulk spin-flop transition.  For $H' < H < H_{SF}$ there is no surface
spin-flop phase in the strict sense; instead, the surface restructures
by, in effect, producing a discommensuration infinitely far away in
the bulk.  The results are used to explain in detail the phase
transitions occurring in systems consisting of a finite, even number
of layers.
\end{abstract}
%]

%\narrowtext
\section{Introduction}
\label{sec:intro}

	It has been known for a long time that if an antiferromagnet
with suitable anisotropy is placed in an external magnetic field $H$
parallel to the easy axis (the axis along which the spins are aligned,
in opposite directions on different sublattices, in zero magnetic
field) and the field strength is increased, a first-order transition
will occur \cite{Neel} in which the spins are realigned in directions
(approximately) perpendicular to the applied field, but with a
component along the field direction.  The transition to this {\it spin
flop} phase occurs when $H$ is equal to a spin-flop field $H_{SF}$,
whose value depends on the exchange energy and the anisotropy. As $H$
continues to increase beyond $H_{SF}$, the spins on the two
sublattices rotate towards the field direction till eventually, if the
field is sufficiently large, they are parallel to each other in a
ferromagnet structure.

	 In 1968 Mills \cite{M68} proposed that in an antiferromagnet
with a free surface, spins near the surface could rotate into a
flopped state at a field $H'_{SF}$ significantly less than $H_{SF}$.
This {\it surface spin-flop} (SSF) problem was later studied by Keffer
and Chow \cite{KC}, who found a transition at $H'_{SF}$, but to a
state having a character rather different than that proposed by Mills.
Interest in this problem was recently rekindled through experimental
work on layered structures consisting of Fe/Cr(211)
su\-per\-lat\-ti\-ces\cite{WM94,WM94b}.  If the thickness of the Cr
layers is chosen appropriately, adjacent Fe blocks are coupled
antiferromagnetically, and thus in zero magnetic field they exhibit an
antiferromagnetic structure in which the magnetization of each layer
is opposite to that of the adjoining layers.  Applying an external
magnetic field parallel to the layers can give rise to phase
transitions in which the magnetization in certain layers rotates or
reverses its direction, and the results found experimentally depend
upon whether the number of Fe layers is even or odd.  The experimental
work has motivated a number of theoretical and numerical studies of
finite and semi-infinite systems \cite{WM94,WM94b,T94,T95,Pap2,T98}.
Most of these have found evidence for the existence of SSF states.

	In the present paper we address the issue of the existence of
SSF phases and some related topics by studying the properties of the
ground states of chains of antiferromagnetically coupled classical
$XY$ spins, each spin variable represented by an angle $\theta$
between 0 and $2\pi$, subject to a uniaxial anisotropy $D$ as well as
to an external magnetic field $H$, as a function of $D$ and $H$.  One
can think of $\theta$ as the direction of the magnetization in an Fe
layer in a superlattice, or of the average magnetization in a layer of
an antiferromagnet containing spins belonging to one type of
sublattice.  Minimizing the energy of a one-dimensional model then
corresponds to minimizing the free energy of a three-dimensional
layered system, provided fluctuations inside the layers do not have a
drastic effect. This means that the model we consider here is, in its
essentials, equivalent to those used in previous studies.  It allows
us to come to some fairly definite conclusions about SSF phases in
semi-infinite systems, and about the behavior of systems containing a
finite number of layers.  Our principal conclusions were published
previously in a short report\cite{MG97}; the present paper contains
the complete argument, and supplies a number of additional details.

	In order to understand the properties of finite and
semi-infinite chains, it is helpful to begin with an infinite chain
and a defect structure known as a ``discommensuration'' (or
``soliton'' or ``kink''), which can occur in both the
antiferromagnetic and the spin-flop phases.  In Sec.~\ref{sec:infch}
we work out the properties of the discommensurations of minimum energy
in the antiferromagnetic ground state of the $XY$ chain. Using these
results, we obtain, in Sec.~\ref{sec:seminfch}, a phase diagram for
surface phase transitions in a semi-infinite chain.  Both
discommensurations and surface phase transitions are essential for
understanding the properties of finite chains.  These are discussed in
Sec.~\ref{sec:finch}, where we provide a comprehensive and detailed
explanation of the complicated series of transitions found in chains
containing an even number of spins.

	The numerical procedures we used to study the phase diagram
are described in Secs.~\ref{sec:infch} and \ref{sec:seminfch}, and a
certain number of analytic results are derived in
Sec.~\ref{sec:anres}.  The concluding Sec.~\ref{sec:concl} provides a
summary, and notes some topics which still need to be studied.

\section{Infinite Chain}
\label{sec:infch}

	We consider an infinite chain of classical $XY$ spins described
by the Hamiltonian
\begin{equation} 
{\cal H}  = \sum_{i=-\infty}^\infty\biggl\{ \cos(\theta_i
- \theta_{i+1}) - H \cos \theta_i +  {D \over 4} [ 1 - \cos (2 \theta_i)]
\biggr\},
\label{eqn:AFhaminf}
\end{equation}
where the antiferromagnetic exchange coefficient has been taken as the
unit of energy, $\theta_i$ is the angle between the direction of the
$i$th spin and the external magnetic field $H$, and $D$ is a two-fold
spin anisotropy.
	Our aim is to identify the zero-temperature phases of this
system, that is, those which minimize the energy.  Minimizing the
energy of a one-dimensional system corresponds to minimizing the free
energy of a layered three-dimensional system when the fluctuations
within each individual layer are not playing an important role, as is
the case for the Fe/Cr superlattices mentioned in
Sec.~\ref{sec:intro}.

\begin{figure}[htbp]
\centerline{\psfig{figure=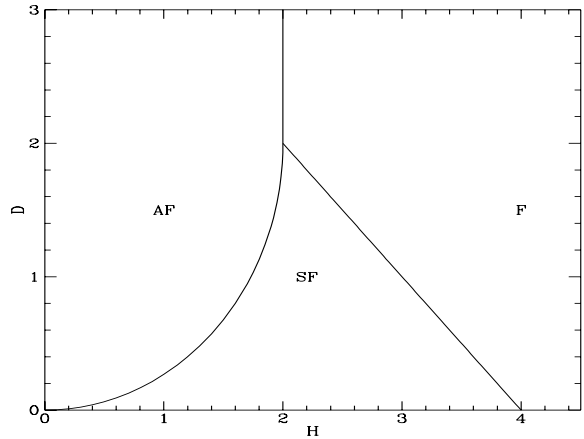,width=3.35in,height=2.5in}}
\caption{Phase diagram for an infinite chain. The AF, F and SF
regions are occupied by the antiferromagnetic, ferromagnetic and spin-flop
phases respectively.}
\label{fig:AFinf}
\end{figure}

The phase diagram of the system consists of three separate regions, as
shown in Fig.~\ref{fig:AFinf}. For $D >2$, the line $H=2$ separates
the ferromagnetic (F) configuration, with all the spins parallel to
the field, from the antiferromagnetic (AF) one with the spins
alternating between 0 and $\pi$, parallel and anti-parallel to the
field.  Along the AF:F boundary the ground state is infinitely
degenerate since it is possible to flip any number of non-adjacent
spins in the F chain with no change in energy.

	For $D<2$ and intermediate values of $H$, the ground state no
longer corresponds to spins in the Ising positions, $\theta_i$ equal 0
or $\pi$, but is a spin-flop (SF) phase in which the spins alternate
between $+ \phi$ and $-\phi$, where
\begin{equation}
\cos \phi  = H/(4-D)\ .
\end{equation}
The spin-flop region extends between the boundaries
$H=\sqrt{D(4-D)}$ and $H=4-D$ which are first and second order,
respectively \cite{AC}.

We now consider the case when an infinite chain is constrained by
suitable boundary conditions to include a discommensuration (for
detailed studies of discommensurations in Frenkel-Kontorova models
see, for example, references [\onlinecite{A81,A83a,A83b,RBG90}]).  The
study of the discommensuration phase diagram is important because it
helps to understand the minimal energy configurations observed both in
semi-infinite and finite systems.
	A discommensuration is a defect which can arise in a periodic
phase whose period is two or greater.  In particular, the AF ground
state has period two and is degenerate: for one ground state
$\theta_i=0$ for $i$ even and $\pi$ for $i$ odd; for the other,
$\theta_i=\pi$ for $i$ even and $0$ for $i$ odd.  A discommensuration
results if one requires that a single configuration $\{\theta_i\}$
approach one of these ground states as $i$ tends to $-\infty$ and the
other as $i$ tends to $+\infty$; for instance,
\begin{eqnarray}
  \theta_{2n} \to 0,\ \theta_{2n+1} \to \pi &\mbox{ as }& n\to +\infty
\nonumber\\
  \theta_{2n} \to \pi,\ \theta_{2n+1} \to 0 &\mbox{ as }& n\to -\infty.
\label{}
\end{eqnarray}

	The defect energy of a discommensuration is the difference
between the energy of the configuration containing the
discommensuration and the energy of the corresponding ground
state. Since both of these energies are infinite for an infinite
chain, a proper definition requires some care; see, e.g.,
[\onlinecite{TG88}].  We are interested in discommensurations which,
for a given $D$ and $H$, minimize this energy; they constitute what we
call the discommensuration phase diagram.  It is convenient to start
by considering the limiting case $D=\infty$, where the spins are
constrained to lie along the Ising positions.  For $0 < H <2$ the
discommensuration of minimum energy is a configuration in which two
successive spins someplace in the middle of the chain are parallel to
the field $H$:
\begin{equation}
\dots,0,\pi,0,\pi,0,0,\pi,0,\pi,\dots\ .
\label{eqn:Isdisc}
\end{equation}
In the following we will use the notation AF$^\prime$ to label this
phase. When $H=2$, due to the absence of further-than-nearest-neighbor
interactions, there is not a unique minimum-energy discommensuration
associated with the AF phase.  One can have any arbitrary even number
of spins aligned with the field, not just two, as in
(\ref{eqn:Isdisc}), and other, more complicated defects are possible.
The ferromagnetic ground state for $H\geq 2$ has period one and is
non-degenerate, so there are no discommensurations.

	As the spin anisotropy $D$ decreases from infinity, lower
energies may occur if in a discommensuration the spins are not limited
to the Ising values $0$ and $\pi$.  For these cases it is difficult to
find an explicit analytic form for the minimum energy
discommensuration, and one has to use numerical techniques to tackle
the problem.
The numerical procedure that we have adopted relies on the method of
effective potentials [\onlinecite{CG,KH}], which is very efficient for
obtaining the ground state of models with short range interactions and
discretized variables.  The main advantage of this method is that it
yields the true ground state, rather than some metastable one.  The
main disadvantage for our problem is that it requires the spin
variables to be discretized: they can take on only a finite number of
values.  We generally used a discretization grid in which each
$\theta_i$ is an integer times $2 \pi / 1400$.  To overcome the
effects of the discretization we first fixed the anisotropy at some
intermediate value, typically $D \approx 0.6$, then used the
Chou-Griffiths algorithm\cite{CG} to identify minimal energy states of
different phases for the system of discretized spins, and, finally,
employed the equilibrium equations,
\begin{equation}
{\partial {\cal  H} \over \partial \theta_i} =0,
\label{eqn:eqeqns}
\end{equation}
for continuous spins in order to refine the configurations obtained
using discretized spins.  The phase boundaries located by comparing
the energies of neighboring phases, calculated using the refined
configurations, were then followed as the value of $D$ was changed in
small steps, while the spin configurations were updated using
(\ref{eqn:eqeqns}). The location of the phase boundaries was then
checked against those obtained starting with finer discretization
grids. We established that, using a discretization of $2 \pi /1400$,
the error in the location of the boundary, $\Delta H$, was in the
range of $10^{-8}-10^{-9}$ throughout the range of $D$ values we
studied.
	The procedure just described was used to find minimum energy
configurations of a ring of spins (periodic boundary conditions) of
length $L$ with $L$ odd, so as to produce a configuration containing a
discommensuration.  When $L$ is large (we used $L \le 31$) compared to
the size of the discommensuration, this is practically the same as
studying the minimal energy discommensuration in an infinite chain.

\begin{figure}[htbp]
\centerline{\psfig{figure=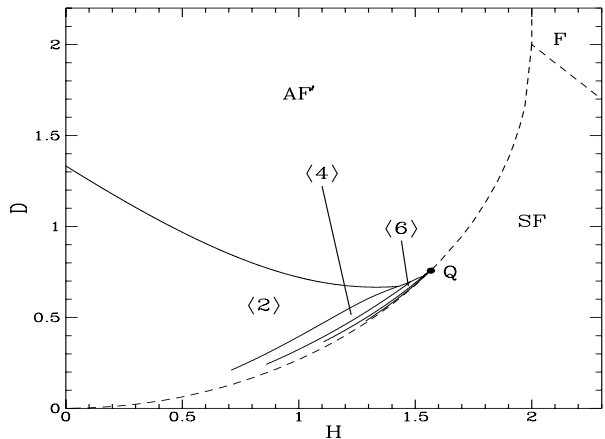,height=2.5in,width=3.35in}}
\caption{Discommensuration phase diagram for an infinite chain. The
dashed phase boundaries correspond to phase transitions in the
discommensuration-free chain, the solid lines in 
Fig.~\protect{\ref{fig:AFinf}}.}
\label{fig:AFpd}
\end{figure}

 The numerical results are summarized in the discommensuration phase
diagram in Fig.~\ref{fig:AFpd}.  There are, of course, no
discommensurations in the F phase.  As for the SF phase, our numerical
results showed a smooth variation of spin angles with $D$ and $H$, and
consequently no phase transitions.  However, various phase transitions
were identified for AF phase discommensurations.  In the AF$^\prime$
region, Fig.~\ref{fig:AFpd}, the spins in the minimum energy
discommensuration stay locked in their $D=\infty$ positions. The
persistence of this Ising spin locking for finite values of the
anisotropy is a rather common feature in models with two-fold spin
anisotropy [\onlinecite{MY94,HMY95,HMY95b}]. Here it has the
consequence that the multiphase degeneracy encountered at the point
$(H=2,D=\infty)$ persists throughout the locus $(D\ge 2, H=2)$.

\begin{figure}[htbp]
\centerline{\psfig{figure=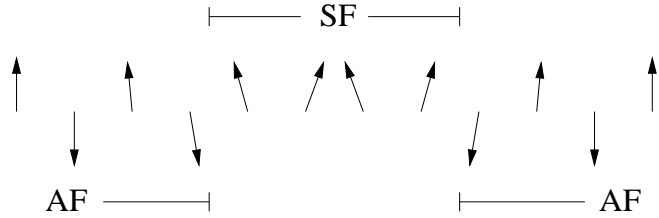,width=3.35in}}
\caption{Schematic representation of phase $\langle 4 \rangle$ for
moderate values of the spin anisotropy. The
phase can be regarded as resulting from merging a portion of the
spin-flop phase (SF) with two semi-infinite antiferromagnetic chains
(AF). The spins nearest the AF-SF and SF-AF interfaces are expected
to relax from their ideal AF or SF angles.}
\label{fig:AFsfins}
\end{figure}

	For values of $D$ lying below the lower boundary of
AF$^\prime$, but still inside the AF region in Fig.~\ref{fig:AFpd},
``flopped'' discommensurations of different length have lower energies
than the Ising discommensuration (\ref{eqn:Isdisc}).  A flopped
discommensuration of type $\langle 2m\rangle$ consists of a ``core''
of $2m$ spins in which the spin configuration resembles that in a bulk
spin-flop phase, located between ``tails'', each of which rapidly
reverts to the configuration of the corresponding AF phase with
increasing distance from the core (see Fig.~\ref{fig:AFsfins}).  One
can think of the region where the core changes into the tail as an
``interface'' between the AF phase out in the tail and the SF phase in
the core.  From this perspective, the discommensuration consists of a
pair of interfaces, AF-SF and SF-AF, bounding the SF core.  As $D$
decreases, these interfaces broaden, making the distinction between
the ``tails'' and the ``core'' less clear, but we continue to use the
same label $\langle 2m\rangle$ for the discommensuration which evolves
continuously from the one with a clearly-defined core of size $2m$ at
larger $D$.

\begin{figure}[htbp]
\centerline{\psfig{figure=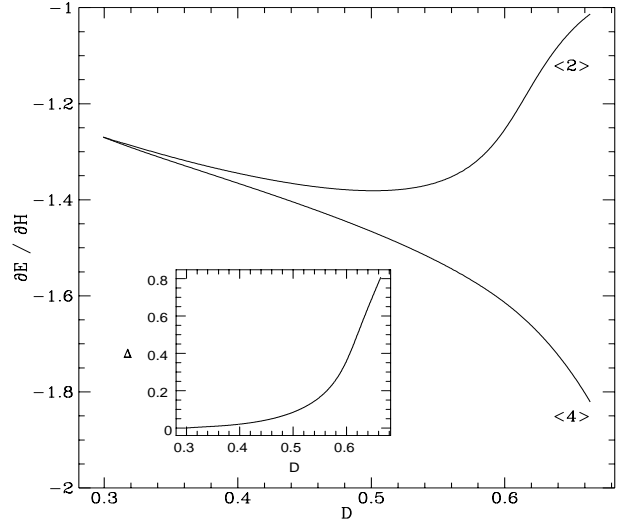,width=3.35in,height=3.0in}}
\caption{Plot of the derivative of the energy with respect to field in the 
two neighboring phases, $\langle 2 \rangle$, $\langle 4 \rangle$ along their
common boundary, for a ring of 17 spins.  The inset shows the difference
$\Delta$ of the two derivatives.}
\label{fig:AFderiv}
\end{figure}

	An analytic calculation, see Sec.~\ref{sec:anres}, shows that
the equation for the second-order transition between AF$^\prime$ and
$\langle 2 \rangle$ in Fig.~\ref{fig:AFpd} is
\begin{equation}
(D+H -1)^{-1} = 5/3 + D -H,
\label{e.boundary1}
\end{equation}
\noindent in good agreement with our numerical calculations, and those in
[\onlinecite{Pap}] when $H=0$.
	At low values of $H$, the discommensuration $\langle 2
\rangle$ has the lowest energy, but upon approaching the bulk AF:SF
phase boundary, one finds a sequence of phase transitions to $\langle
4 \rangle$, $\langle 6 \rangle$, \dots \ as $H$ increases, as shown in
Fig~\ref{fig:AFpd}.  Our numerical procedures found values of $2m$ up
to 14, and we were able to trace the first-order lines separating the
different $\langle 2m \rangle$ phases down to a value of $D$ between
0.1 and 0.4.
For smaller values of $D$, the difference $\Delta$ of the energy
derivatives $\partial E / \partial H$ in two neighboring phases was no
longer sufficient to allow us to distinguish the phases numerically
and locate the phase boundary.  See the example in
Fig.~\ref{fig:AFderiv}.  However, we found no evidence that these
lines terminate in critical points.  The smooth decrease of $\Delta$
shown in the inset of Fig.~\ref{fig:AFderiv} contrasts with what one
might expect at a critical point (as in Fig.~\ref{fig:AFendpoint1}).
Therefore, it seems plausible to assume that the $\langle 2m \rangle :
\langle 2m +2 \rangle$ boundaries persist all the way down to $D=0$.

	The sequence of transitions associated with a broadening of
the discommensuration can be understood in the following way. The
defect energy of a discommensuration can be thought of as the sum of
the energy required to produce a pair of AF-SF and SF-AF interfaces
infinitely far apart, an interaction energy between the interfaces
which we assume is positive and rises rapidly as they approach each
other, and a ``bulk'' contribution proportional to the size of the
core, arising from the fact that in the AF part of the phase diagram,
the SF phase is metastable.  In terms of which discommensuration has
the lowest energy, the interface repulsion obviously favors a large
core, and the metastable ``penalty'' a small core.  The actual size
will represent some compromise between the two.  Upon approaching the
AF:SF boundary, the metastable penalty goes to zero, so the
discommensuration of minimum energy should become larger and larger.
Hence one expects the $\langle 2m\rangle : \langle 2m+2 \rangle$
boundary to approach the AF:SF transition line as $m \to \infty$.
This is consistent with our numerical calculations, and in agreement
with the predictions of Papanicolaou [\onlinecite{Pap}]. (Note,
however, that these transitions reflect the discrete nature of the
spin chain and therefore are absent in the continuum approximation
employed in [\onlinecite{Pap}] for small spin anisotropy.)
	The triple points at which the phases AF$^\prime$, $\langle 2m
\rangle$ and $\langle 2m+2 \rangle$ meet tend to an accumulation
point, $Q$, located at $H\approx 1.58, D\approx 0.78$.  This should be
the point at which the energy to create a pair of AF-SF and SF-AF
interfaces infinitely far apart is equal to the energy of an Ising
discommensuration.

\section{Semi-infinite chains}
\label{sec:seminfch}

	We now consider the surface states of a semi-infinite
chain. The Hamiltonian for the system is the same as
(\ref{eqn:AFhaminf}) but with the sum extending only over non-negative
values of $i$ ($i=0$ denotes the surface site):

\begin{equation} 
{\cal H}  = \sum_{i=0}^\infty\biggl\{ \cos(\theta_i
- \theta_{i+1}) - H \cos \theta_i +  {D \over 4} [ 1 - \cos (2 \theta_i)]
\biggr\}\ .
\label{eqn:AFhamseminf}
\end{equation}

\begin{figure}[htbp]
\centerline{\psfig{figure=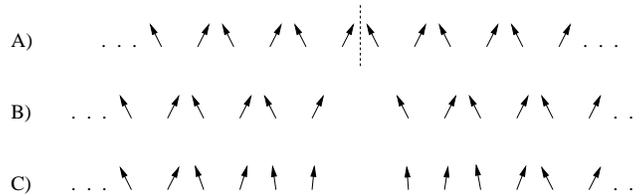,width=3.35in}}
\caption{Cutting an infinite chain in two (a) while keeping the spins
``frozen'' results in two semi-infinite chains with unreconstructed surfaces
(b). Allowing the spins to relax to positions which minimize the energy
typically results in reconstruction of the surface (c), a re-arrangement of the
spins nearest the surface.}
\label{fig:cut}
\end{figure}

	It is useful to think of semi-infinite chains as obtained by
cutting an infinite chain in two. Removing a bond in the infinite
chain without allowing the spins to move will give two semi-infinite
chains that we shall term unreconstructed. If the spins of the
unreconstructed chains are then allowed to relax, a rearrangement of
the spins near the surface may take place, as illustrated in
Fig.~\ref{fig:cut}, which lowers the energy.  Notice that even though
the total energy of the semi-infinite chain is infinite, {\it changes}
in the energy when a configuration is modified near the surface (or in
a way such that the modifications decrease sufficiently rapidly with
increasing distance from the surface) are well defined.  We want to
consider surface states which minimize the energy in the sense that no
local modifications of the configuration near the surface can decrease
the energy.

	The task of finding the reconstructed surface of minimum
energy is, in general, not simple (except when all the spins in the
chain are subject to the Ising locking). To identify the minimal
energy surface states we used numerical algorithms based on effective
potential methods that, as mentioned earlier, require a discretization
of the spin variables at each site.  It is important to notice that,
since the $\theta_i$'s are constrained to take on only discrete
values, after a finite distance, or ``penetration depth'' $l$ from the
surface the spins will be {\em exactly}\/ in the discretized positions
corresponding to a doubly-infinite chain or an unreconstructed
surface.  Configurations for the infinite chain were obtained using
the Floria-Griffiths algorithm \cite{FG} which, within the limits of
the discretization, yields the exact ground state.
	Next, the Chou-Griffiths algorithm \cite{CG} with its
successive iterations was used to generate reconstructed surface
configurations $\{ \theta_0, \theta_1, ... ,\theta_l\}$.  This should
give the exact configuration minimizing the surface energy for the
discrete spins.  However, in practice we had to limit $l$ to a maximum
value $l_{max}$ no larger than 50; thus the method could not yield the
correct configuration for a larger penetration depth.  The phase
boundaries were then identified as explained in the previous section.

	The resulting phase diagram is shown in
Fig.~\ref{fig:AFseminf}.  Throughout the F region the minimum energy
surface states are simply the unreconstructed surfaces; it is easy to
see that making any changes will increase the energy. In the SF
region, since the ground state of the infinite chain has period two,
there are two unreconstructed surfaces.  Each of them undergoes a
reconstruction in which the spins nearest the surface tilt towards the
magnetic field direction, as in Fig.~\ref{fig:cut}(c). However, this
change in spin direction occurs continuously as a function of $H$ and
$D$, and so no surface phase transitions are observed inside the SF
region.

	Next consider the AF part of the phase diagram.  Again there
exist two possible surface states, $A$ and $B$, whose unreconstructed
versions, $A_u$ and $B_u$, have surface spins parallel ($\theta_0 =0$)
or opposite ($\theta_0 = \pi$) to the field direction:
\begin{eqnarray}
A_u &=& \{ 0, \pi, 0 ,\pi,0, \pi, ... \}, \label{eqn:AFconf1}\\
B_u &=& \{\pi, 0 ,\pi,0,\pi,0, ... \}\label{eqn:AFconf2}.
\end{eqnarray}
A surface will be said to be of type $A$ ($B$) if the spin
configuration tends to that of $A_u$ ($B_u$) far from the surface.

	Throughout the AF region of the phase diagram, the minimum
energy surface of type $A$ is the unreconstructed $A_u$.  However, the
$B$-type surface shows a number of different structures in different
parts of the AF region, as indicated in Figs.~\ref{fig:AFseminf} and
\ref{fig:AFseminf2}.  In region AF$_1$ the unreconstructed surface
$B_u$ has the lowest energy. In region AF$_2$, which meets AF$_1$
along a line $H=1$ for $D$ larger than the value at $O$, it is
energetically favorable to flip the surface spin so that it points
along the field direction, and there is a set of degenerate (equal
minimum energy) reconstructed surfaces
\begin{equation}
[ 0 \rangle =0, 0, \pi, 0, \pi, 0, \pi \ldots,\quad
[ 2 \rangle =0, \pi, 0, 0, \pi, 0, \pi \ldots,
\label{e.surf}
\end{equation}
and so forth, where $[2n\rangle$ consists of $2n$ spins $0,\pi,\ldots$
in an antiferromagnetic arrangement, followed by two spins parallel to
the field, and then the bulk antiferromagnetic phase.  One can think
of this reconstructed surface as an Ising discommensuration, whose
core consists of two adjacent spins with $\theta_i=0$, located a
distance $2n$ from the surface.  Because the ``tails'' of this
discommensuration have zero length, it does not interact with the
surface, and its energy is independent of its distance from the
surface.  While this degeneracy persists throughout the AF$_2$ region,
along the line $D\ge 2,H=2$ the degeneracy is even greater: the set of
minimum energy surface states includes cases where the number of
consecutive spins pointing along the field is not limited to 2 but can
attain any even number, e.g., $\{ 0,0,0,0,\pi,0,\pi...\}$ or
$\{0,\pi,0,\pi,0,0,0,0,0,0,\pi,0, \pi...\}$.  Incidentally, we note
that these degeneracies are somewhat artificial in that they would be
lifted by introducing weak longer-range interactions in the
Hamiltonian (\ref{eqn:AFhamseminf}).

	In the AF$_3$ region of Fig.~\ref{fig:AFseminf} the $B$-type
surface again reconstructs, but the spin anisotropy is sufficiently
low that the spins unlock from the Ising angles.  As in the AF$_2$
region, one can think of the surface state as consisting of a
discommensuration located a finite distance from the surface, but now
this discommensuration is of the flopped type with a core of length
two, and tails extending out on either side of the core.  We again
employ the notation $[2n\rangle$ for the surface state with $2n$ spins
to the left of the core, that is, in the tail extending to the
surface.  Because of this tail, the discommensuration interacts with
the surface, and the minimum surface energy occurs for a specific
value of $2n$, depending upon $D$ and $H$.  Thus, in AF$_3$, one finds
genuine spin-flop surface states.  As $H$ increases, the
discommensuration moves further from the surface. It does this, at
least when $D$ is large, discontinuously in steps of 2, via a series
of first-order phase transitions, some of which are shown in
Fig.~\ref{fig:AFseminf2}, where they extend leftwards from the point
$P$.  For smaller values of $D$, the edges of the core are not as well
defined, and it is more difficult to associate the $[ 2n \rangle \to [
2n +2 \rangle$ transitions with a discontinuous jump of the
discommensuration.  Numerically we have seen states with $2n$ up to
14, and our results are consistent with $n$ tending to infinity at the
right side of the AF$_3$ region, which our analytic calculations
(Sec.~\ref{sec:anres}), in agreement with [\onlinecite{T94}], show to
be the line
\begin{equation}
 D=\sqrt{1+H^2} -1.
\label{e.bound1}
\end{equation}
The upper boundary of the AF$_3$ region extending from $O$ to $P$ is a
continuous (second-order) transition.  One can think of it as the
limit of stability of the Ising surface phase $[0\rangle$ as $D$
decreases inside AF$_2$. An analytic calculation,
Sec.~\ref{sec:anres}, shows that the implicit equation for the
boundary is:
\begin{eqnarray}
 && ( 2 + D - H -1/a)^{-1}= 2 + D +H -a,
 \nonumber\\
 && a:= H + D + 1/(1 -H -D).
\label{e.bound2}
\end{eqnarray}
Thus the point $P$, where all the phases
$[2n\rangle$ come together, lies at $H=4/3,$ $D=2/3$, the intersection of
(\ref{e.bound1}) and (\ref{e.bound2}).  Both (\ref{e.bound1}) and
(\ref{e.bound2}) agree with our numerical results.

	We find that the first-order lines extending downwards and
leftwards from $P$ in Fig.~\ref{fig:AFseminf2}, separating phases
$[2n\rangle$ from $[2n+2\rangle$, end in critical points as $D$
decreases. This is clearly visible in the example in
Fig.~\ref{fig:AFendpoint1}, which shows the typical behavior of the
energy derivatives of two neighboring phases along their coexistence
line. Near a critical point $D=D_c$ one expects $\Delta$ to vary as
$\sqrt{D-D_c}$, in qualitative agreement with what we observed.
	The larger the value of $n$, the further the first-order line
extends towards the origin of the $H,D$ plane, but presumably for any
finite value of $n$ the difference between the phases $[2n\rangle$ and
$[2n+2\rangle$ eventually disappears at some finite value of
$D$. Because this value decreases with increasing $n$, it is plausible
that the corresponding critical points accumulate at the origin.

	As is evident in Fig.~\ref{fig:AFseminf}, the region AF$_3$
becomes extremely narrow as $D$ decreases. The left boundary
approaches a parabola $D=0.5 H^2$ to within numerical precision, which
is asymptotically the same as (\ref{e.bound1}).  We nonetheless
believe that the width of AF$_3$ remains finite as long as $D>0$.
Numerical evidence for this is shown in Fig.~\ref{f3} where the value
of the surface spin, $\theta_0$, at the left edge of the AF$_3$ region
(that is for $H$ just large enough to produce the surface spin-flop
phase) is plotted as a function of $D$. The results are for
$l_{max}=34$ spins in the surface layer (see the description of the
numerical approach given above).  Below $D=0.05$ the results become
unreliable because $l_{max}$ is too small, as we can tell by carrying
out calculations for different values of $l_{max}$.  However,
extrapolating from larger values of $D$ indicates that as $D$ goes to
zero, $\theta_0$ tends to a value near $\pi/3$ or $60^\circ $, showing
that even for very small $D$ the discommensuration at the threshold
field is still a finite distance from the surface.  This situation is
quite distinct from that in region AF$_1$, where $\theta_0=\pi$, and
in AF$_4$, discussed below, where $\theta_0=0$.

	Between AF$_3$ and the AF:SF bulk phase boundary lies region
AF$_4$, see Figs.~\ref{fig:AFseminf} and \ref{fig:AFseminf2}, in which
the flopped discommensuration is repelled by the surface, so that its
minimum energy location is in the bulk infinitely far away from the
surface, as noted in \cite{T94}. Thus there is no minimum-energy
reconstructed $B$ surface, or, properly speaking, a ``surface
spin-flop phase'' in region AF$_4$.  It seems better to identify
AF$_4$, thought of as part of the $B$-type surface phase diagram, as a
``discommensuration phase'', since the minimum energy surface will
always be of the $A$-type, with the surface spin $\theta_0=0$.

	In Fig.~\ref{fig:fig10} the discommensuration phase diagram
for the infinite chain (Fig.~\ref{fig:AFpd}), represented by dashed
lines, is superimposed on the $B$-type surface diagram for the
semi-infinite chain, represented by solid lines, in the vicinity of
points $P$ and $Q$, which are common to both diagrams, as is the
broken line (shown dashed) from $P$ to $Q$.  Note that the $OP$ line
of the surface diagram, Fig.~\ref{fig:AFseminf2}, lies above the lower
boundary of the AF' region of the discommensuration phase diagram in
Fig.~\ref{fig:AFpd}.  Thus to the left of $P$, for $H < 4/3$, as $D$
decreases the reconstructed $B$-type surface phase changes from Ising
to a flopped form before the corresponding change is energetically
favorable for the bulk discommensuration.

	In addition, Fig.~\ref{fig:fig10} shows that the part of the
$H,D$ plane corresponding to $\langle 2m\rangle$ in the
discommensuration phase diagram, Fig.~\ref{fig:AFpd}, for $2m\geq 4$
lies entirely inside the AF$_4$ region of Fig.~\ref{fig:AFseminf} (and
\ref{fig:AFseminf2}) for the surface phase diagram.  This is
consistent with our observation that as long as the discommensuration
is a finite distance from the surface, in the AF$_3$ region, it is
always of the type $2m=2$.  Thus as $H$ increases, it is only {\it
after} the discommensuration has moved infinitely far from the
surface, and thus has no influence on the surface phase diagram, that
its core begins to broaden.

      In retrospect it seems likely that the broadening of the SSF
transition mentioned in the abstract of [\onlinecite{KC}] actually
refers to broadening of the bulk discommensuration which, as noted
above, occurs as $H$ approaches the AF:SF boundary inside region
AF$_4$.  It appears that no work prior to ours has correctly
identified the stable SSF phase at small values of $D$, characterized
when it first appears with increasing $H$ by a surface spin with a
value very near $60^\circ $ (Fig.~\ref{f3}).  The narrowness of the
AF$_3$ region for small $D$ may be why it was overlooked.

\section{Finite Chain}
\label{sec:finch}

	We now move on to consider the case of a chain of finite
length $L$. Since a surface reconstruction can occur at both ends of
the chain, and it is also possible for a discommensuration to be
present in the interior of the chain, we write its total energy in the
form
\begin{equation}
E_L = L \epsilon + E_s^L + E_s^R + E_d,
\label{eqn:AFminim}
\end{equation}
where $\epsilon$ is the bulk energy, the ground-state energy per spin
for an infinite chain, $E_s^L$ and $E_s^R$ are the energies of the
left and right surfaces respectively, and $E_d$ is the energy of a
discommensuration in the chain (if present). Minimizing the total
energy for fixed $L$ is equivalent to finding the spin configuration
that minimizes $E_s^L + E_s^R + E_d$.

	In writing (\ref{eqn:AFminim}), $L$ was assumed to be
sufficiently large that the interaction between the two ends of the
chain, and between each end and the discommensuration, if present, can
be neglected. For any given $L$ this condition can always be satisfied
by choosing a large enough value for the spin anisotropy.  Outside the
range of $D$ for which (\ref{eqn:AFminim}) holds, the behavior of the
system will depend strongly on the actual length of the chain. Since
we are not interested in $L$-dependent features of the phase diagram,
apart from whether $L$ is even or odd, we shall assume that $L$ is
sufficiently large to justify the use of (\ref{eqn:AFminim}).

	From the discussion presented in the previous sections one can
predict that a finite chain will not undergo any phase transition for
values of $D$ and $H$ inside the SF and F regions.  On the other hand
it can also be anticipated that the behavior of the chain in the AF
region will be rather complicated.  As noted in
[\onlinecite{WM94,WM94b,T95}], the behavior of the chain for values of
$D$ and $H$ in the AF region changes dramatically according to whether
the length of the chain is even or odd.

	If $L$ is odd, both ends of the chain have to be of the same
type, $A$ or $B$, unless a discommensuration is present.  Having two
$A$-type surfaces gives a lower energy than two $B$-type surfaces,
because the former results in a net magnetization in the direction of
the field, and the latter a net magnetization opposite to the
field. Similar considerations show that throughout the AF region it is
energetically unfavorable to insert a dislocation, thus producing one
$A$-type and one $B$-type surface.  Hence for odd $L$, the minimum
energy corresponds to two (unreconstructed) $A$-type surfaces at
either end of the chain, and no discommensurations.

	On the other hand, when $L$ is even, the two surfaces have to
be of different types, unless a discommensuration is present.  The
analysis of Sec.~\ref{sec:infch} has shown that discommensurations are
not favored energetically outside region AF$_4$.  Thus, for $D$ and
$H$ falling in region AF$_1$ or AF$_3$, one expects one surface of
type $A$ and the other of type $B$.  Moreover, from the results of
Sec.~\ref{sec:seminfch}, we expect that in region AF$_1$ the $B$
surface remains unreconstructed, whereas surface spin-flop states
should be observed in AF$_3$ owing to the reconstruction of the
$B$-type end of the chain.  The $A$-type end of the chain remains, of
course, in its unreconstructed state.
	Next, in region AF$_4$ the energy is minimized using two
$A$-type surfaces and a discommensuration, which lies at the center of
the finite chain because it is repelled by both surfaces.  Finally, in
AF$_2$, because of the degeneracy due to the Ising spin locking, one
has either a reconstructed $B$-type surface or a discommensuration,
depending upon what one wants to call it, and an $A$-type surface at
the other end of the chain.

	Consequently, if $D$ is smaller than the value corresponding
to point $P$ in Fig.~\ref{fig:AFseminf2}, we expect a finite system
with even $L$ to undergo the following set of transitions with
increasing $H$.  At $H=0$, Fig.~\ref{fig:AFfinitetr}(a), there are
unreconstructed surfaces of types $A$ and $B$ at opposite ends of the
chain.  When $H$ reaches the threshold for the formation of an SSF
phase, the $B$-type surface restructures discontinuously, (b) to form
a type $\langle 2 \rangle$ discommensuration which then, as $H$
increases, moves towards the center of the chain in a series of
discontinuous steps, (c) and (d), some of which may be continuous if
$D$ is smaller than the value for the corresponding critical point,
see Sec.~\ref{sec:seminfch}.

	The discommensuration will reach the center of the chain,
Fig.~\ref{fig:AFfinitetr}(d), when $H$ is close to the threshold for
the AF$_4$ or discommensuration region in Fig.~\ref{fig:AFseminf}.
Further increases of $H$ will lead to a broadening of the
discommensuration, with $\langle 2m \rangle$ going through the
sequence $\langle 2 \rangle, \langle 4 \rangle, \langle 6 \rangle,
\ldots$ of Fig.~\ref{fig:AFpd}; see Fig.~\ref{fig:AFfinitetr}(d) to
(g).  While these transitions are likely to be discontinuous for
larger values of $D$, it may be hard to see the discontinuities when
$D$ is small.  The center of the $\langle 2m \rangle$
discommensuration in Fig.~\ref{fig:AFfinitetr} does not fall at the
precise center of the chain when $m$ is even; the offset is needed so
that the surface spins can both be (approximately) parallel, rather
than antiparallel, to the field direction.  (For $L=12$ the offset
occurs when $m$ is odd.)

	The AF-SF and SF-AF interfaces on either side of the core move
outwards as the discommensuration expands, and eventually they reach
the surfaces of the chain, Fig.~\ref{fig:AFfinitetr}(g), at a field
very close to that required to produce the bulk spin-flop transition.
At still higher fields the entire chain can be thought of as being in
the bulk spin-flop phase, with appropriate (reconstructed) surface
configurations corresponding to this phase.  Sufficiently large values
of $H$ will eventually force all of the spins into the ferromagnetic
configuration $\theta_i=0$.

	The scenario just described is basically consistent with
previous numerical studies, including two that have appeared quite
recently \cite{Pap2,T98}, and our own numerical work.  Thus
Fig.~\ref{fig:AFplochi1} shows the magnetic susceptibility $\chi
=\partial M/\partial H$, $M$ the magnetization, for a chain of $L=22$
spins when $D=0.5$.  The spikes appearing in Fig.~\ref{fig:AFplochi1}
should be Dirac delta functions. Here they appear to have a finite
height because of the finite incremental step $\delta H$ chosen for
the numerical calculation.
The first spike in Fig.~\ref{fig:AFplochi1} (for $H \approx 0.9$)
signals the transition from the AF$_1$ region into the surface
spin-flop AF$_2$, phase $[0 \rangle$. The first series of spikes, for
$H$ between 0.9 and 1.13, is associated with first-order spin-flop
transitions, in agreement with \cite{T95,T98}. For $H$ between 1.13
and 1.32, one observes a second series of transitions associated with
the broadening of the discommensuration.
Figure \ref{fig:AFchi2} shows the susceptibility for the same length
of chain ($L=22$) with a smaller anisotropy, $D=0.3$.  The spikes are
smaller than in Fig.~\ref{fig:AFplochi1} due to decrease in
anisotropy, and some of the surface spin-flop peaks have disappeared,
which is what one would expect in view of the critical points along
the $[ 2n \rangle: [2n+2 \rangle$ phase boundaries noted in
Sec.~\ref{sec:seminfch}.

	A recent study by Papanicolaou \cite{Pap2} of the dynamics of
a model similar to (\ref{eqn:AFhaminf}), but with three-dimensional
(classical) spins, shows evidence for metastability and hysteresis as
the magnetic field $H$ is varied, as one would expect for a
first-order SSF transition.  Additional hysteresis is seen as the
field is increased beyond the SSF transition, consistent with
additional first-order transitions of the sort discussed above.  Small
differences in detail between these results and ours can probably be
explained in terms of hysteresis effects, or possibly as due to the
fact that the models are not identical.  A numerical study of
(\ref{eqn:AFhaminf}) by Trallori \cite{T98}, using an area-preserving
map, is also in very good agreement with all of our results, except
that certain transitions which we would expect to be first order as
the discommensuration moves to the center of the chain and broadens
are found to be continuous when $D$ is very small.  But this
difference is probably not important, since the discontinuities will
in any case be very small when $D$ is small, and could be absent
because $L$ is finite.

\section{Analytical results}
\label{sec:anres}

In this last section we give a detailed derivation of the analytical
results presented earlier in the paper.  As already noted, analytical
solutions to the problem of minimizing the energy are, in general,
only available when the spins are in Ising position, $\theta=0$ or
$\pi$.  However, when deviations from these values are small,
systematic approximations are possible.  Throughout this section we
shall use $\theta^0_i$ to indicate Ising or ``locked'' spin values,
$\theta_i$ for the actual canted values, and $\tilde{\theta}_i \equiv
\theta_i - \theta^0_i$ for the deviations of the latter from the
locked values.

	To obtain an analytic expression for a second-order boundary
separating locked and canted versions of a spin configuration, we
start by expanding (\ref{eqn:eqeqns}) to first order in the spin
deviations, assuming that they are small,
\begin{equation}
\cos(\theta^0_i - \theta^0_{i-1}) ( {\tilde{\theta}}_{i}- {
{\tilde{\theta}}_{i-1}} ) + \cos(\theta^0_{i+1} - \theta^0_{i}) (
{\tilde{\theta}}_{i}- { {\tilde{\theta}}_{i+1}} ) = (H
\cos(\theta^0_i) + D) {\tilde{\theta}}_{i}\ ,
\label{eqn:AFlinrec0}
\end{equation}
and then solving these equations self-consistently.

	We first apply this strategy to find the boundary separating
phases AF$^\prime$ and $\langle 2 \rangle$, Fig.~\ref{fig:AFpd}, using
the labels for sites in the flopped discommensuration $\langle 2
\rangle$ given in Fig.~\ref{fig:AFinfch1}.  The equations
(\ref{eqn:AFlinrec0}) can be written as recursion relations, in terms
of ratios $x_i = \tilde{\theta}_i / \tilde{\theta}_{i-1}$ of
successive spin deviations, in the form:
\begin{eqnarray}
{x_{2j}^{-1}} + x_{2j+1} &=& 2 + D +H \mbox{ for }j \le -1\ ,
  \nonumber\\
{x_{2j+1}^{-1}} + x_{2j+2} &=& 2 + D -H\mbox{ for }j \le -1\ ,
  \nonumber\\
{x_{0}^{-1}} - x_{1} &=& D +H \ , 
  \nonumber\\
-{x_{1}^{-1}} + x_{2} &=& D +H \ , 
  \nonumber\\
{x_{2j}^{-1}} + x_{2j+1} &=& 2 + D -H \mbox{ for }j \ge 1\ ,
  \nonumber\\
{x_{2j+1}^{-1}} + x_{2j+2} &=& 2 + D +H\mbox{ for }j \ge 1,
\label{eqn:AFf}
\end{eqnarray}
with a solution
\begin{eqnarray}
{x_{2j}} &=& {s_2}\mbox{ for }j \ge 1, 
  \nonumber\\
{x_{2j+1}} &=& {s_1}\mbox{ for }j \ge 1. 
  \nonumber\\
{x_{0}^{-1}} - x_{1} &=& D +H,
  \nonumber\\
-{x_{1}^{-1}} + x_{2} &=& D +H,
  \nonumber\\
x_{2j+2} &=& {s_2^{-1}}\mbox{ for }j \le -1, 
  \nonumber\\
x_{2j+1}&=& {s_1^{-1}}\mbox{ for }j \le -1. 
\label{eqn:AFx2j}
\end{eqnarray}
obtained using techniques of continued fractions. Here $s_1$ and $s_2$ are
given by
\begin{eqnarray}
s_1 &=& 2 [2+D-H]\, [ (2+D+H) (2+D-H) + t ]^{-1}\ , 
  \nonumber\\
s_2 &=& (1/2)\, [2+D+H + t/(2+D-H)] \ ,
 \nonumber\\
t &:=& \sqrt{ (2+D+H)^2 (2+D-H)^2 - 4 (2+D+H) (2+D-H) }.
\label{eqn:AFs1s2}
\end{eqnarray}
The only set of values $(H,D)$ for which equations (\ref{eqn:AFx2j})
can be simultaneously satisfied under the constraint that the modulus
of $s_1$ and $s_2$ cannot exceed 1 (so that the spin deviations decay
to zero infinitely far from the discommensuration core) has to satisfy
the relation
\begin{equation}
  (D +H -1)^{-1} = 5/3 + D - H \ ,
\label{eqn:AFboundary1}
\end{equation}
which is the same as (\ref{e.boundary1}). Equation
(\ref{eqn:AFboundary1}) identifies the locus of points where the spin
deviations for phase $\langle 2 \rangle$ become vanishingly small,
which is the second-order boundary AF$^\prime: \langle 2 \rangle$.

	The same method can be used to find the second-order boundary
$OP$ between AF$_2$ and AF$_3$ in Fig.~\ref{fig:AFseminf} or
\ref{fig:AFseminf2}. In the $[ 0\rangle$ phase close to the border,
with the spins labeled as in Fig.~\ref{fig:AFch1-2}, deviations from
the corresponding Ising configuration, (\ref{e.surf}), will be small,
and the solution to (\ref{eqn:AFlinrec0}) takes the form
\begin{eqnarray}
{x_{2j}} &=& {s_2}\mbox{ for }j \ge 1\ ,
 \nonumber\\
{x_{2j+1}} &=& {s_1}\mbox{ for } j\ge 1\ ,
 \nonumber\\
 x_1 &=& 1 - H -D \ \ ,
 \nonumber\\
 - {x_1^{-1}} + s_2 &=& H+D \ ,
\label{eqn:AFA}
\end{eqnarray}
using the same notation introduced previously, with $s_1$ and $s_2$ again
defined by  (\ref{eqn:AFs1s2}). 
These equations yield an additional 
relation for $s_2$,
\begin{equation}
s_2 = H+D + [ 1 - H - D]^{-1},
\label{eqn:AFstar}
\end{equation}
 which can be satisfied together with (\ref{eqn:AFs1s2})
only on the locus of points $\Gamma$ defined by equation
(\ref{e.bound2}).

	A similar analysis assuming small deviations from Ising values
for the state $[2 \rangle$ shows that the point $P$ on $\Gamma$,
Fig.~\ref{fig:AFseminf2}, occurs at the intersection of the curve
\begin{equation}
1+D+H = (1+D-H)^{-1},
\label{eqn:AFconstraint}
\end{equation}
with the boundary (\ref{eqn:AFboundary1}), so that $P$ falls at
$H=4/3, D=2/3$, in good agreement with our numerical results
$H=1.333,\ D=0.6666$.  Likewise, one can show that the other $[2n
\rangle$ states for $n > 1$ meet the AF$_2$ region at $P$, which is a
sort of multicritical point for the surface phase diagram.

	A somewhat different approach yields an equation for the
boundary between the AF$_3$ and AF$_4$ regions, that is, the left edge
of the AF$_4$ region in Figs.~\ref{fig:AFseminf} and
\ref{fig:AFseminf2}.  As this corresponds to an accumulation of
surface spin-flop states $[2n\rangle$ as $n\to\infty$, the distance
from the surface of the chain to the core of the dislocation will
become arbitrarily large, so that the spin angles in the
discommensuration are essentially independent of distance from the
surface\cite{MY94,SY}, as confirmed by our numerical calculations.
Hence, by a route analogous to that described in
[\onlinecite{MY94,SY}], it is possible to evaluate the energy
difference between two neighboring phases, $\Delta E_n = E_{[
2n\rangle} - E_{[2n+2 \rangle}$, by iterating the equilibrium
equations (\ref{eqn:eqeqns}) on either side of the discommensuration.

	Using the fact that the spin deviations at the surface are
becoming vanishingly small, one obtains, to leading order at large
$n$,
\begin{equation}
\Delta E_n \approx {1 \over 2} (\tilde{\theta}_1- \tilde{\theta}_0)^2
+ {1 \over 2} (\tilde{\theta}_2- \tilde{\theta}_1)^2
+ {1 \over 2} D (\tilde{\theta}_1^2 + \tilde{\theta}_0^2) 
-{1 \over 2} H (\tilde{\theta}_1^2 - \tilde{\theta}_0^2) \ ,
\label{eqn:AFed}
\end{equation}
where the $\tilde{\theta}_i$'s are the spin deviations of phase $[2n+2
\rangle$. The expression for $\Delta E_n$ can be simplified by using
(\ref{eqn:AFlinrec0}) to express $\tilde{\theta}_1$ and
$\tilde{\theta}_2$ in terms of $\tilde{\theta}_0$, noting that when
$i=0$, the term $\cos(\theta^0_i - \theta^0_{i-1}) (
{\tilde{\theta}}_{i}- { {\tilde{\theta}}_{i-1}} )$ must be omitted
from (\ref{eqn:AFlinrec0}), because $i=0$ represents the left edge of
the finite chain, (\ref{eqn:AFhamseminf}).  Substituting
\begin{eqnarray}
\tilde{\theta}_1&\approx&(1 +D +H) \tilde{\theta}_0 \ , \nonumber \\
\tilde{\theta}_2&\approx&[2 + D - H - (1 +D +H)^{-1}]
\tilde{\theta}_1\ ,
\label{eqn:AFstst}
\end{eqnarray}
 into (\ref{eqn:AFed}) gives
\begin{eqnarray}
  &&\Delta E_n= {\textstyle {1\over2}} W 
\tilde{\theta}_0^2 + {\cal O}({\tilde{\theta}}_0^3)\ ,
\nonumber\\
 W&:=& 2D + 7{D^2} + 5{D^3} + {D^4} +(2D+ {D^2}) H
\nonumber\\
	&&-(1 + 5 D + 2D^2)H^2 - {H^3} + {H^4}.
\label{eqn:AFdebig}
\end{eqnarray}
Note that this expression holds for all values of $D$ as long as
$\tilde{\theta}_0$ is small, that is, the discommensuration is very
far from the surface.  But this means that an accumulation of states
$[2n\rangle$ as $n$ tends to infinity must lie on a locus where $W$ in
(\ref{eqn:AFdebig}) vanishes, because in region AF$_3$ the
discommensuration is attracted by the surface ($W>0$), while it is
repelled in AF$_4$ $(W<0)$.  The relevant root of this equation takes
the simple form
\begin{equation}
D = \sqrt{1 + H^2} -1\ ,
\label{eqn:AFbssf}
\end{equation} 
in agreement with [\onlinecite{T94}], and with our numerical calculations.

\section{Conclusions}
\label{sec:concl}

\begin{figure}[htbp]
\centerline{\psfig{figure=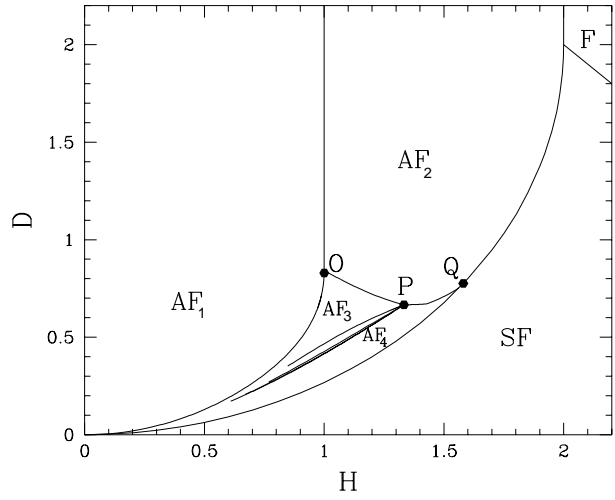,width=3.35in}}
\caption{Phase diagram for a semi-infinite chain with a $B$-type
surface. More details of the AF$_3$ region are visible in 
Fig.~\protect{\ref{fig:AFseminf2}}.}
\label{fig:AFseminf}
\end{figure}

\begin{figure}[htbp]
\centerline{\psfig{figure=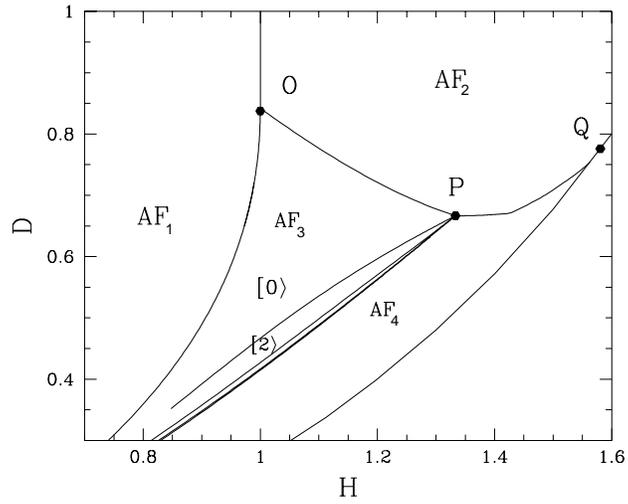,width=3.35in}}
\caption{Detail of the phase diagram for a semi-infinite chain with a
$B$-type surface.}
\label{fig:AFseminf2}
\end{figure}

\begin{figure}[htbp]
\centerline{\psfig{figure=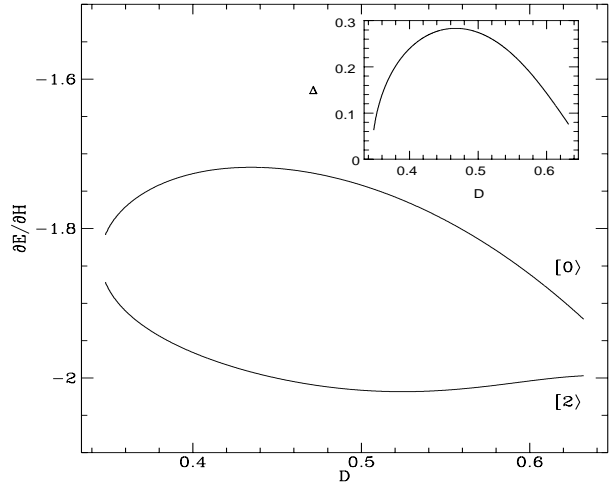,width=3.35in,height=2.8in}}
\caption{Plot of the derivative of the energy with respect to field in the 
two neighboring phases $[0\rangle$ and $[2\rangle$ along their common boundary,
using 50 spins in the surface layer.  The inset shows the difference $\Delta$
of the two derivatives.}
\label{fig:AFendpoint1}
\end{figure}

	Our work shows that the structure of surface spin flop (SSF)
states and their relationship to the behavior of finite systems is
significantly more complex than anticipated in previous work.  In
particular, the genuine SSF phase for a semi-infinite system, which we
identify with region AF$_3$ in our surface phase diagram,
Figs.~\ref{fig:AFseminf} and \ref{fig:AFseminf2}, has previously been
confused with what we call the ``discommensuration'' phase, region
AF$_4$, in which the $B$-type surface has, strictly speaking,
completely disappeared through a restructuring in which a
discommensuration has moved infinitely far away from the surface into
the bulk.  The fact that both the SSF and the discommensuration phase
occur at a magnetic field $H$ significantly below that required to
produce a bulk spin flop transition, together with the extremely small
interval of $H$ over which the SSF phase is stable when the anisotropy
$D$ is small, are no doubt the reason the two have not been
distinguished in previous studies.  Nonetheless, they are quite
different phenomena, and distinguishing them is essential for a proper
understanding of phase transitions associated with surfaces, both in
semi-infinite and finite systems.

	Our results for the discommensuration and surface phase
diagrams lead to very definite and detailed predictions, discussed in
Sec.~\ref{sec:finch}, for the complicated sequence of phase
transitions occurring in a system with an even number of layers
(spins) as $H$ increases at fixed $D$.  They are in good agreement
with various numerical studies, including our own, if allowance is
made for the uncertainties inherent in numerical work of this sort,
and this gives us additional confidence in the validity of our
analysis.  To the extent that this model antiferromagnet correctly
describes Fe/Cr superlattices, we can also claim to have achieved a
basic understanding of the processes giving rise to the phase
transitions observed experimentally in the latter.

	That does not, of course, mean that our model is adequate for
understanding SSF phases and other surface phase transitions in more
traditional antiferromagnets, such as MnF$_2$.  However, as noted in
Sec.~\ref{sec:intro}, minimizing the energy of a one-dimensional model
is the analog of minimizing the free energy of a three-dimensional
layered system, whenever each layer can be described, using mean-field
theory or in a purely phenomenological way, by means of a total
magnetization serving as a sort of order parameter.  To be sure, the
parameters which enter the Hamiltonian for the one-dimensional chain
may not be those appropriate for three-dimensional system.  But one
can still expect qualitative similarities in the phase diagrams, even
if certain quantitative aspects are different.

	In that connection, it is appropriate to ask whether certain
features of the discommensuration and surface phase diagrams of the
one-dimensional model depend in a sensitive way upon the particular
form of the Hamiltonian (\ref{eqn:AFhaminf}).  For example, it
contains no spin coupling beyond nearest neighbors, whereas it would
be physically more realistic to assume, at the very least, some sort
of exchange coupling of further neighbors, decreasing rapidly with
distance.  Would introducing such interactions lead to significant
changes in the phase diagram?  Could they, for example, make the SSF
phase disappear entirely at low values of the anisotropy?

\begin{figure}
\centerline{\psfig{figure=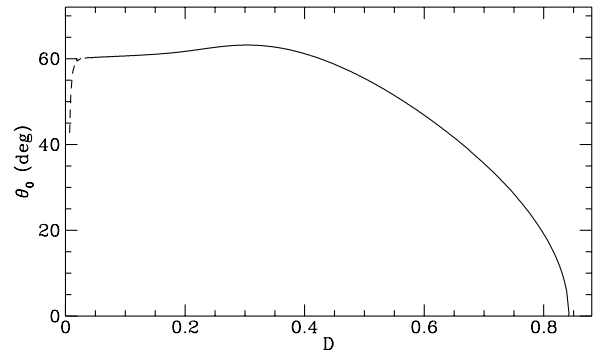,width=3.35in}}
\caption{Surface spin $\theta_0$ along the left edge of the AF$_3$
region as a function of anisotropy $D$. The surface layer consisted of
34 spins, and the behavior of the curve at low $D$ (dashed) is affected by
finite-size effects in the numerical calculations.  }
\label{f3}
\end{figure}

\begin{figure}[htbp]
\centerline{\psfig{figure=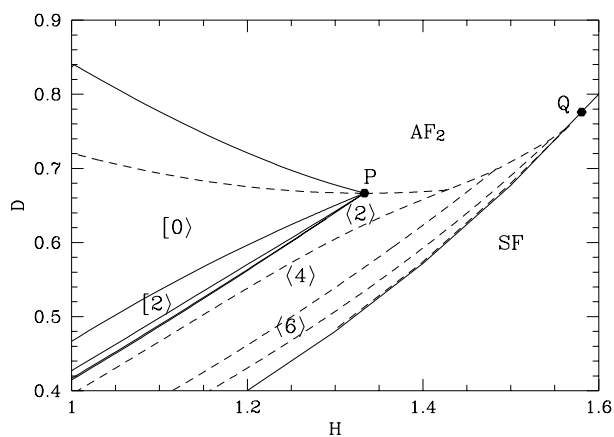,width=3.35in}}
\caption{%
Discommensuration phase diagram (Fig.~\protect{\ref{fig:AFpd}}), using dashed
lines, superimposed on the phase diagram for a semi-infinite chain with a
$B$-type surface (Fig.~\protect{\ref{fig:AFseminf2}}), using solid lines, in
the vicinity of the point $P$. The broken line connecting $P$ with $Q$ is part
of both phase diagrams.}
\label{fig:fig10}
\end{figure}

\begin{figure}
\centerline{\psfig{figure=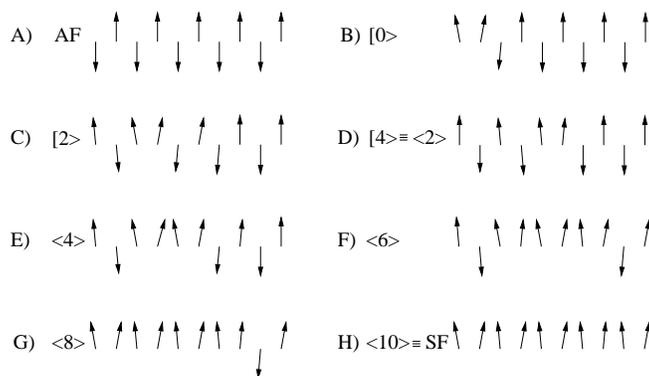,width=3.35in}}
\caption{Schematic representation of the series of different phases
encountered in a chain of 10 spins for increasing values of $H$.}
\label{fig:AFfinitetr}
\end{figure}

\begin{figure}
\centerline{\psfig{figure=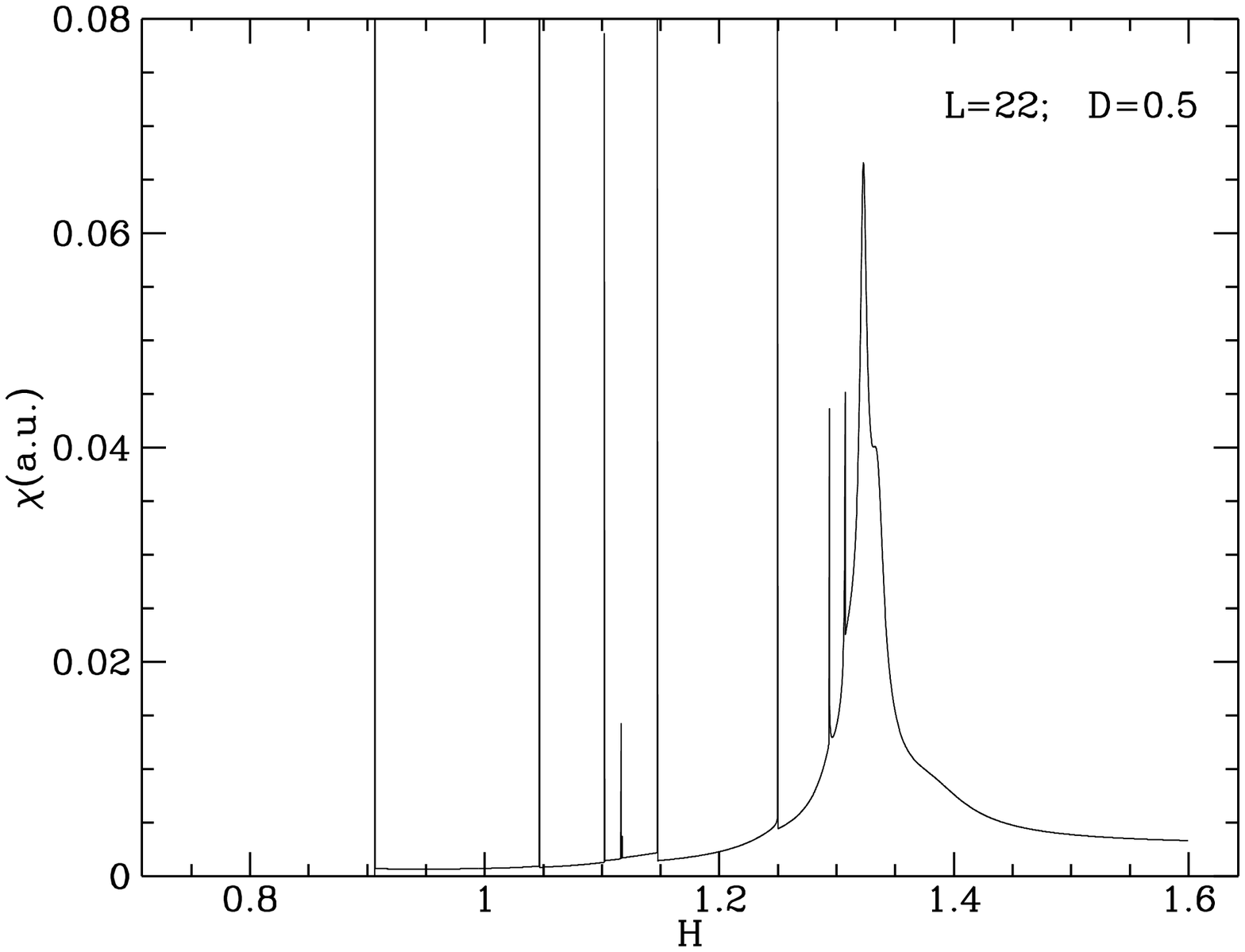,width=3.35in}}
\caption{Plot of the susceptibility (in arbitrary units) for a chain
of 22 spins for $D=0.5$.}
\label{fig:AFplochi1}
\end{figure}

\begin{figure}
\centerline{\psfig{figure=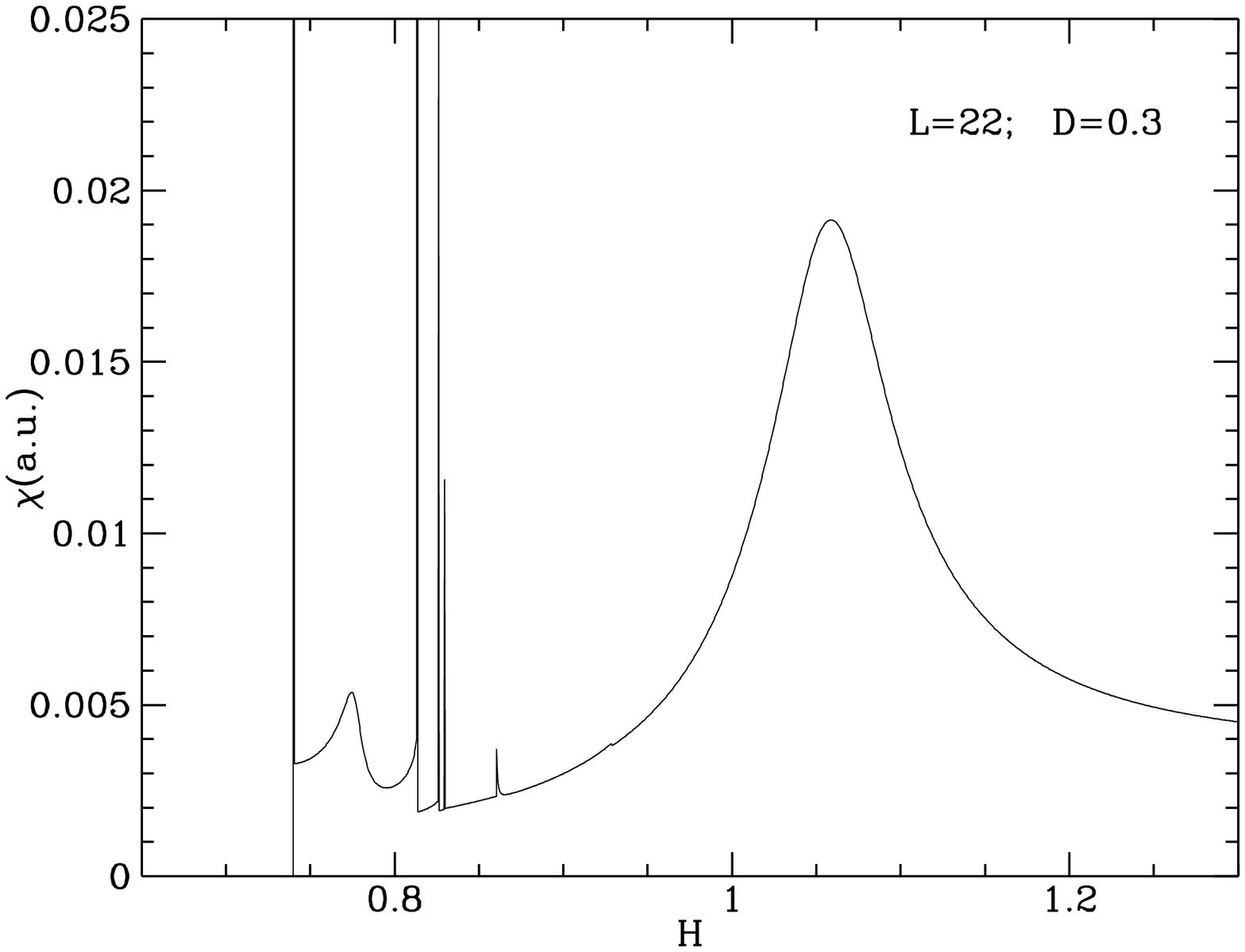,width=3.35in}}
\caption{Plot of the susceptibility (in arbitrary units) for a chain
of 22 spins for $D=0.3$.}
\label{fig:AFchi2}
\end{figure}

\begin{figure}[htbp]
\centerline{\psfig{figure=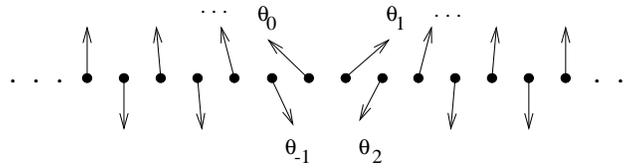,width=3.35in}}
\caption{Schematic representation of the canted discommensuration phase 
$\langle 2\rangle$.}
\label{fig:AFinfch1}
\end{figure}

\begin{figure}[htbp]
\centerline{\psfig{figure=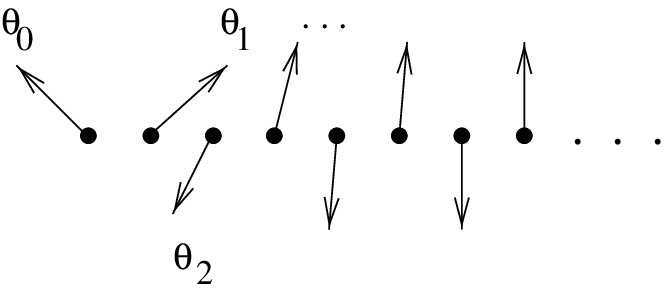,width=3.35in}}
\caption{Schematic representation of the surface phase $[0\rangle$.}
\label{fig:AFch1-2}
\end{figure}

	This is one of many questions which cannot be answered
definitively in advance of appropriate calculations.  It is worth
pointing out that our physical picture of the SSF phase as due to a
discommensuration finding its minimum energy at a finite distance from
the surface does not seem to depend on the absence of further-neighbor
exchange (or possibly other types of) interaction, so we can well
imagine that the phenomenon persists with a more realistic
Hamiltonian.  Nonetheless, this is one respect in which our work
remains incomplete. While our numerical results, especially the
apparent existence of a non-zero limit for $\theta_0$ as $D$ goes to
zero, Fig.~\ref{f3}, support our description in terms of a
discommensuration, an appropriate analytic calculation in the limit of
small $D$, of the sort which might (among other things) give the value
of this limiting angle, has not been carried out. Such a study would
probably provide insight into whether weak further-neighbor
interactions simply change the quantitative values of various
parameters, or lead to a qualitatively different result, such as the
absence of the AF$_3$ region when $D$ is sufficiently small.

	It seems unlikely that weak further-neighbor interactions
would remove the first-order transitions between the surface phases
$[2n\rangle$ and $[2n+2\rangle$, or change the fact that these
transitions terminate in critical points as $D$ decreases. On the
other hand, such a modification of the Hamiltonian would surely remove
the degeneracy of the surface states in the AF$_2$ region of
Figs.~\ref{fig:AFseminf} and \ref{fig:AFseminf2}.  Thus one would not
be surprised to find significant modifications in the phase diagram
near the multicritical point $P$.  Indeed, $P$ which might well
disappear, to be replaced by some other, more complicated, structure
allowing the different $[2n\rangle$ phases to disappear as $H$
increases.  Also, sufficiently strong further-neighbor interactions of
the proper kind might result in the infinite-chain discommensurations
undergoing their broadening transitions at significantly smaller
values of the magnetic field $H$. This could lead to a complicated
surface phase diagram in which the minimum energy discommensurations
broaden while they are still a finite distance from the surface.  How
this might effect the $[2n\rangle$ to $[2n+2\rangle$ transitions and
their critical points is hard to guess in advance of actually doing a
calculation.

	Hence there is much which remains to be understood about
surface spin-flop transitions in antiferromagnets.  Nonetheless, we
believe that the calculations, numerical and analytical, presented in
this paper have served to sort out some important physical effects,
and in this sense our results provide a solid foundation for future
work.

%Fig. 1

%Fig. 2

%Fig. 3

%Fig. 4

%Fig. 5

%Fig. 6

%Fig. 7

%Fig. 8

%Fig. 9

%Fig. 10

%Fig. 11

%Fig. 12

%Fig. 13

%Fig. 14

%Fig. 15


\begin{thebibliography}{11}
%\begin{references}

\bibitem{Neel} 
L. N\'eel, {\em Ann. Phys.} (Paris) {\bf 5}, 232 (1936)

\bibitem{M68}
D. L. Mills, Phys. Rev. Lett. {\bf 20}, 18 (1968).

\bibitem{KC} 
F. Keffer and H. Chow, Phys. Rev. Lett. {\bf 31}, 1061 (1973).

\bibitem{WM94} R. W. Wang, D. L. Mills, E. E. Fullerton, J. E. Mattson
and S. D. Bader, Phys. Rev. Lett. {\bf 72}, 920 (1994).


\bibitem{WM94b} R. W. Wang and D. L. Mills, Phys. Rev. B {\bf 50} 3931 (1994).

\bibitem{T94}
L. Trallori, P. Politi, A. Rettori, M. G. Pini and J. Villain, Phys. Rev. Lett.
{\bf 72}, 1925 (1994).

\bibitem{T95} 
L. Trallori, P. Politi, A. Rettori, M. G. Pini and J. Villain, J. Phys. C {\bf
7}, L451 (1995).

\bibitem{Pap2}
N. Papanicolaou, J. Phys. Cond. Mat {\bf 10}, L131 (1998).

\bibitem{T98}
L. Trallori, Phys. Rev. B {\bf 57}, 5923 (1998).

\bibitem{MG97}
C. Micheletti, R. B. Griffiths and J. M. Yeomans, J. Phys. A {\bf 30}, L233
(1997).

\bibitem{AC} F. B. Anderson and H. B. Callen, Phys. Rev. A, {\bf 136},
1068 (1964)

\bibitem{A81}
S. Aubry, in {\it Solitons and Condensed Matter Physics}, edited by A. R.
Bishop and T. Schneider (Springer Verlag, Berlin, 1981).

\bibitem{A83a}
S. Aubry, J.  Phys. (Paris) {\bf 44}, 147 (1983).

\bibitem{A83b}
S. Aubry, Physica D {\bf 7}, 240 (1983).

\bibitem{RBG90}
R. B.  Griffiths, in {\it Fundamental Problems in Statistical Mechanics VII},
edited by H. van Beijeren (Elsevier, Amsterdam, 1990).

\bibitem{TG88}
L. H. Tang and R. B. Griffiths, J.~Stat.\ Phys.\ {\bf 53} 853 (1988).

\bibitem{CG} 
W. Chou and R. B. Griffiths, Phys. Rev. B {\bf 34}, 6219 (1986).

\bibitem{KH}
K. Hood, J. Comput. Phys. {\bf 89}, 187 (1990).


\bibitem{MY94}
C. Micheletti and J. M. Yeomans, Europhys. Lett. {\bf 28}, 465 (1994).

\bibitem{HMY95} 
A.B. Harris, C. Micheletti and J. M. Yeomans, Phys. Rev. Lett.  {\bf 74}, 3045
(1995).

\bibitem{HMY95b}
A.B. Harris, C. Micheletti and J. M. Yeomans, Phys. Rev. B {\bf 52},
6684 (1995). 

\bibitem{Pap}
N. Papanicolaou, Phys. Rev. B {\bf 51}, 15062 (1995).

\bibitem{FG} 
L. M. Floria and R. B. Griffiths, Numer. Math. {\bf 55}, 565 (1989).


\bibitem{SY}
 F. Seno and J. M. Yeomans. Phys. Rev. B {\bf 50}, 10385 (1994).

%%%%%%%%%%%%%%%%%%%%%%%%%%%%%%%%%%%%%%%%%%%%%%%
%%%%%%%%%%%%%%%%%%%%%%%%%%%%%%%%%%%%%%%%%%%%%%%







%\end{references} 
\end{thebibliography}
\end{document}